\documentclass[journal]{IEEEtran}
\usepackage{graphicx}
\usepackage{amssymb}
\usepackage{amsmath}
\usepackage{xcolor}
\usepackage{bm}
\usepackage{float}
\usepackage{makecell}
\usepackage{url}
\usepackage{hyperref}
\usepackage{threeparttable}
\usepackage{longtable}
\usepackage{titlesec}
\usepackage{nameref}
\usepackage{enumitem}
\usepackage{algorithm}
\usepackage{algpseudocode}
\usepackage{cite}
\usepackage{gensymb}
\usepackage{threeparttable}
\usepackage{epstopdf}

\title{Efficient Sampling and Sensitivity Analysis of Rare Transient Instability Events via Subset Simulation\\
\small
\author{
    \IEEEauthorblockN{
     Jingyu Liu\textsuperscript{1}, 
     Xiaoting Wang\textsuperscript{2},
     Xiaozhe Wang\textsuperscript{1} \\
       }
    \IEEEauthorblockA{
        \textsuperscript{1}\textit{Department of Electrical and Computer Engineering}, \textit{McGill University}, Montréal, Canada\\
        \textsuperscript{2}\textit{Department of Electrical and Computer Engineering}, \textit{University of Alberta}, Edmonton, Canada\\
       jingyu.liu@mail.mcgill.ca;
       xiaotin5@ualberta.ca;
       xiaozhe.wang2@mcgill.ca
    }
    }
\normalsize
\thanks{This work is funded by the Natural Sciences and Engineering Research Council (NSERC) Discovery Grant, NSERC RGPIN-2022-03236. }
}
\date{October 2024}

\begin{document}

\maketitle

\begin{abstract}

    Assessing the risk of low-probability high-impact transient instability (TI) events is crucial for ensuring robust and stable power system operation under high uncertainty. 
    However, direct Monte Carlo (DMC) simulation for rare TI event sampling is computationally intensive. 
    This paper proposes a subset simulation-based method for efficient small TI probability estimation, rare TI events sampling, and subsequent 
    sensitivity analysis. Numerical studies on the modified WSCC 9-bus system demonstrate the efficiency of the proposed method over DMC. Additionally, targeted stability enhancement strategies are designed to eliminate rare TI events and enhance the system's robustness to specific transient faults. 
\end{abstract}


\section{Introduction}
The increasing penetration of volatile loads (e.g., electric vehicles) and renewable energy sources (RES) is introducing growing uncertainties into modern power grids. These uncertainties have led to an uncertain power system transient stability (i.e., there is a non-zero probability that the system suffers transient instability) which requires probabilistic assessment.

A common approach for probabilistic transient stability assessment (PTSA) is time-domain simulation (TDS)-based direct Monte Carlo (DMC) method, which offers high accuracy but is time-consuming, particularly for large systems. To improve efficiency, methods like practical dynamic security regions \cite{yue2020probabilistic} and surrogate models (e.g., polynomial chaos expansion \cite{xu2018propagating} \cite{liu2022sparse} and Kriging \cite{ye2022physics}) have been proposed to enable faster PTSA while maintaining reasonable accuracy.

Most previous PTSA studies focus primarily on the average behaviour of system transient stability (e.g., the probability density of the transient stability index within the high-probability region \cite{ye2022physics}). However,  ensuring power system resilience under high uncertainty requires prioritizing the probabilities in the tail of the transient stability index distribution to improve worst-case performance. 
While previous research  has assessed the risk of rare events in probabilistic power flow studies \cite{tan2024scalable} and power system reliability evaluations\cite{ansari2018hybrid} \cite{hua2014extracting}
, using surrogate model-based methods \cite{tan2024scalable},  cross entropy-based importance sampling \cite{ansari2018hybrid} or subset simulation (SubSim) \cite{hua2014extracting}. 
these studies lack comprehensive sensitivity analysis of input parameter uncertainties. Such analyses are critical for identifying influential parameters in rare TI events and designing effective mitigation measures to enhance the stability and robustness of the system under high uncertainty. 
Furthermore, no existing PTSA study, to our knowledge, explicitly addresses rare transient instability (TI) events or their sensitivity analysis. 

In this paper, we propose a subset simulation-based method for efficiently sampling rare TI events and estimating small TI probabilities. System stability is assessed using a critical clearing time (CCT)-based transient stability margin. Samples with negative margins (i.e., transient instability) are further utilized for sensitivity analysis of uncertain sources and to design targeted stability enhancement strategies. For example, in our simulation study, low power consumption by a specific load is identified as the primary contributor to rare TI events. Increasing its power (e.g., via energy storage) significantly enhances the system's robustness to transient faults. Furthermore, our findings suggest that, given uncertainties on the generation side, the load side may need to respond adaptively to maintain generation-load balance and ensure a sufficient stability margin. 

\vspace{-0.4cm}
\section{Problem Formulation}

The transient dynamics 
of conventional power systems are described by the following 
differential-algebraic equations \cite{liu2022sparse}:
\begin{equation}
\begin{aligned}
\setlength{\abovedisplayskip}{1pt}
\setlength{\belowdisplayskip}{1pt}
\dot{\mathbf{x}}_s &= \bm{f}(\mathbf{x}_s, \mathbf{y}, \bm{\lambda}, \mathbf{u})  \\
\bm{0} &= \bm{h}(\mathbf{x}_s, \mathbf{y}, \bm{\lambda}, \mathbf{u})
\label{eq:transient_dae}    
\end{aligned}
\end{equation}
where $\bm{f}$ and $\bm{h}$ denote the differential (e.g., swing equations) and algebraic equations (e.g., power flow), respectively. $\mathbf{x}_s$ are state variables including generator rotor angles and speeds; $\mathbf{y}$ are algebraic variables (e.g., bus voltages); $\bm{\lambda}$ are system parameters (e.g., load power); $\mathbf{u}$ denote 
discrete events (e.g., fault occurrence and switching operation of circuit breakers).

\begin{figure}[!ht]
\vspace{-0.2cm}
\centering
\includegraphics[width=1\columnwidth]{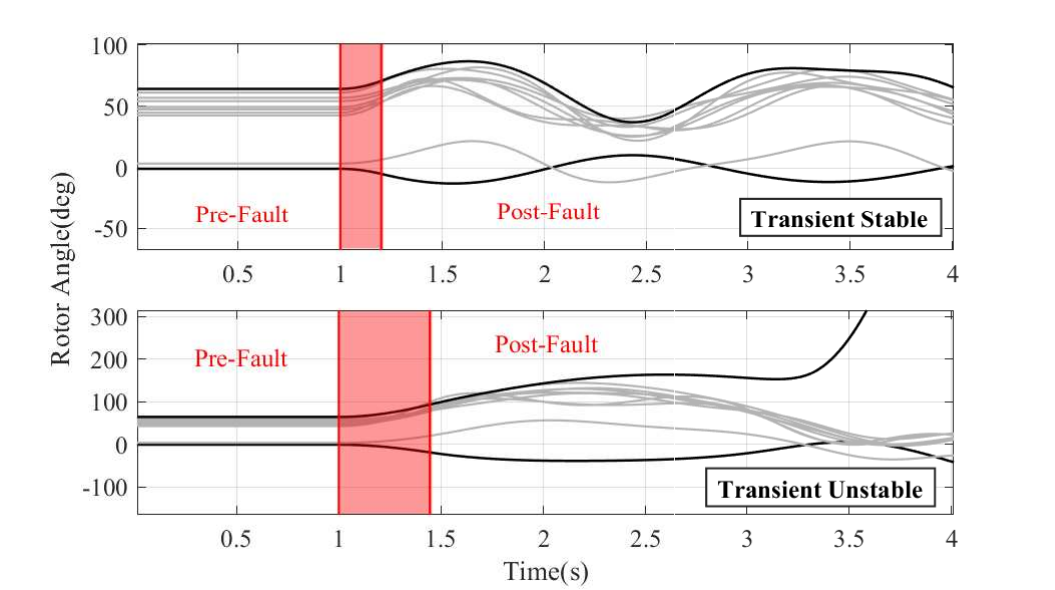}
\vspace{-0.65cm}
\caption{Illustration of transient rotor angle stability and fault clearing time (FCT).  Trajectories exhibiting the largest rotor angle difference are highlighted in black. The fault-on periods with different FCTs are highlighted in red. A large FCT can render the system transient unstable.}
\label{fig:CCT_explain}
\vspace{-0.6cm}
\end{figure}

Fig. \ref{fig:CCT_explain} presents a set of rotor angle trajectories obtained from TDS (i.e., numerical integration) of \eqref{eq:transient_dae}. 
 In this figure, 
 a three-phase ground fault is applied at $t = 1.0$s and cleared after a specified 
 fault clearing time $T_{\mathrm{fct}}$, i.e., marking the fault-on period highlighted in red. The maximum value of $T_{\mathrm{fct}}$ that maintains system stability is the critical clearing time ($T_{\mathrm{cct}}$). The difference  $T_{\mathrm{cct}} - T_{\mathrm{fct}}$ is a well-accepted measure of the transient stability margin (TSM) \cite{liu2022sparse}.

 To model the impact of RES and loads, certain system parameters $\bm{\lambda}$ are treated as random variables $\bm{X}=\{X_i, i=1,...,M\}$ (e.g.,  uncertain sources like wind speeds) 
 whereas the others $\bm{\lambda}^{\prime}$ remain constant. Then, \eqref{eq:transient_dae} is reformulated as:
 \begin{equation}
\begin{aligned}
\setlength{\abovedisplayskip}{2pt}
\setlength{\belowdisplayskip}{2pt}
         \dot{\mathbf{x}}_s &= \bm{f}(\mathbf{x}_s, \mathbf{y}, \bm{\lambda}^{\prime}, \bm{X},  \mathbf{u})  \\
           \bm{0} &= \bm{h}(\mathbf{x}_s, \mathbf{y}, \bm{\lambda}^{\prime}, \bm{X}, \mathbf{u})
\label{eq:Stochastic-PSDAE}        
\end{aligned}
\end{equation}
In this context, the CCT also becomes a random variable (RV) $T_{\mathrm{pcct}}(\bm{X})$ dependent on $\bm{X}$, which naturally leads to the formalization of the probabilistic TSM (PTSM) \cite{liu2022sparse}:
\begin{equation}
\setlength{\abovedisplayskip}{2pt}
\setlength{\belowdisplayskip}{4pt}
     \mathrm{PTSM}:= \mathcal{M}(\bm{X}) = T_{\mathrm{pcct}}(\bm{X}) - T_{\mathrm{fct}}
    \label{eq:PTSM}
\end{equation}
This formulation further introduces a non-zero probability that the system will undergo transient instability:
\begin{equation}
\setlength{\abovedisplayskip}{4pt}
\setlength{\belowdisplayskip}{4pt}
    P_f = \mathbb{P}(\mathcal{M}(\bm{X})\leq 0)=\int_{\mathcal{F}}f_{\bm{X}}(\bm{x})d\bm{x}
    \label{eq:TI_prob}
\end{equation}
where $\bm{X}$ is assumed to follow a joint probability density function
(PDF): $\bm{X}\sim f_{\bm{X}}(\bm{x})$ and $\mathcal{F}=\{\bm{x}: \mathcal{M}(\bm{x})\leq 0\}$ denotes the domain of instability.  In other words, a scenario $\bm{x}$ that results in a PTSM close to $0$  (i.e. CCT is close to FCT) is high-impact, as it renders the system vulnerable in terms of transient stability.

Due to its implementation flexibility, TDS-based DMC is commonly employed to solve \eqref{eq:TI_prob}. This method involves repetitive 
 numerical integration of \eqref{eq:Stochastic-PSDAE} for numerous samples of $\bm{X}$, making it computationally intensive, 
 especially with low TI probabilities $P_f$. When $P_f$ is small, a typical DMC sample set contains only a few TI events, requiring a large number $N$ of  DMC samples 
 to ensure the accuracy in estimating  $P_f$ (e.g., $N\geq 10^4$ if $P_f\approx 0.1\%$) \cite{au2014SubSimBook}. 
 To overcome this, we introduce the subset simulation (SubSim) method which improves the accuracy for small $P_f$ (e.g., $P_f<1\%$) estimation compared to DMC under the same number of TDS runs.
 \vspace{-0.3cm}
\section{Subset Simulation-based Transient Instability Probability Estimation}

Subset Simulation is a reliability analysis method developed to efficiently estimate the probability of rare failure events \cite{au2014SubSimBook}. It outperforms DMC by efficiently sampling the low-probability TI domain $\mathcal{F}$, thus providing a more accurate $P_f$ estimation under the same number of TDS-based evaluations of \eqref{eq:PTSM}. Consider a sequence of threshold values $t_k, k=1,...,q $ with  $t_1>\dots>t_q = 0$, and $\mathcal{F}_k=\{\bm{x}:\mathcal{M}(\bm{x})\leq t_k\}$, we have $\mathcal{F}=\mathcal{F}_q \subset \mathcal{F}_{q-1}\subset\dots \subset \mathcal{F}_{1} $.  The essence of SubSim is to  decompose the low-probability TI events $\mathcal{F}$ into a series of relatively higher-probability conditional events according to the fundamental principle of conditional probability \cite{au2014SubSimBook}: 
%
\begin{equation} 
\setlength{\abovedisplayskip}{4pt}
\setlength{\belowdisplayskip}{4pt}
    P_f=\mathbb{P}(\mathcal{F})=\mathbb{P}(\mathcal{F}_1)\prod_{k=2}^{q}\mathbb{P}(\mathcal{F}_k|\mathcal{F}_{k-1}) \label{eq:SubSim_principle}
\end{equation}

In practice, a fixed probability $p_0<0.5$ is assigned to all intermediate subsets, i.e.,
\begin{equation} 
\setlength{\abovedisplayskip}{6pt}
\setlength{\belowdisplayskip}{6pt}
    p_0=\mathbb{P}(\mathcal{F}_1)=\mathbb{P}(\mathcal{F}_2|\mathcal{F}_1)=\dots=\mathbb{P}(\mathcal{F}_{q-1}|\mathcal{F}_{q-2})
    \label{eq:p0}
\end{equation}
Accordingly, the estimated intermediate threshold $ t_k \approx  \hat{t}_k$ is determined as the $p_0$-percentile of $N_{\mathrm{bat}}$ samples drawn from either $f_{\bm{X}}(\bm{x})$ or the conditional PDF $f_{\bm{X}}(\bm{x}|\mathcal{M}(\bm{x})\leq\hat{t}_{k-1})$. Note that 
 direct Monte Carlo sampling of  $f_{\bm{X}}(\bm{x})$ is sufficient to compute $\hat{t}_1$,  whereas calculating $\hat{t}_2, ...,\hat{t}_{q-1}$ relies on the Markov Chain Monte Carlo (MCMC) method \cite{papaioannou2015mcmc}. This is because an efficient method to generate i.i.d. samples from the conditional PDF $f_{\bm{X}}(\bm{x}|\mathcal{M}(\bm{x})\leq\hat{t}_k)$ is generally unavailable. Consequently, the $p_0$-integrated TI probability estimation is as follows:
\begin{equation}
    P_f= p_0^{q-1}\mathbb{P}(\mathcal{F}_{q}|\mathcal{F}_{q-1})\approx p_0^{q-1}\mathbb{P}(\mathcal{F}_{q}|\mathcal{M}(\bm{x})\leq\hat{t}_{q-1})
    \label{eq:pf_sus_estimation}
\end{equation}
where $\mathbb{P}(\mathcal{F}_{q}|\mathcal{M}(\bm{x})\leq\hat{t}_{q-1})>p_0$ is the SubSim-based estimation of the final level conditional probability $\mathbb{P}(\mathcal{F}_{q}|\mathcal{F}_{q-1})$. Let $\{\bm{x}_{q}^{(n)}, n=1, ...,N_{\mathrm{bat}}\}$ represent the $N_{\mathrm{bat}}$ samples drawn from $f_{\bm{X}}(\bm{x}|\mathcal{M}(\bm{x})\leq\hat{t}_q-1)$. Then, $\mathbb{P}(\mathcal{F}_{q}|\mathcal{M}(\bm{x})\leq\hat{t}_{q-1})$ can be estimated by:
\vspace{-0.1cm}
\begin{equation}
\setlength{\abovedisplayskip}{4pt}
\setlength{\belowdisplayskip}{4pt}
\small
    \mathbb{P}(\mathcal{F}_{q}|\mathcal{M}(\bm{x})\leq\hat{t}_{q-1})\approx \hat{P}_{e}=\frac{1}{N_{\mathrm{bat}}}\sum_{n=1}^{N_{\mathrm{bat}}}I_{m<0}(\mathcal{M}(\bm{x}_{q}^{(n)}))
    \label{eq:last_level_prob_est}
\end{equation}
\normalsize
where $I_{m<0}(m)$ is an indicator function (i.e., $I_{m<0}(m)=1$ when $m<0$. Otherwise, $I_{m<0}(m)=0$).

\begin{figure}[!h]
\centering
\includegraphics[width=1\columnwidth]{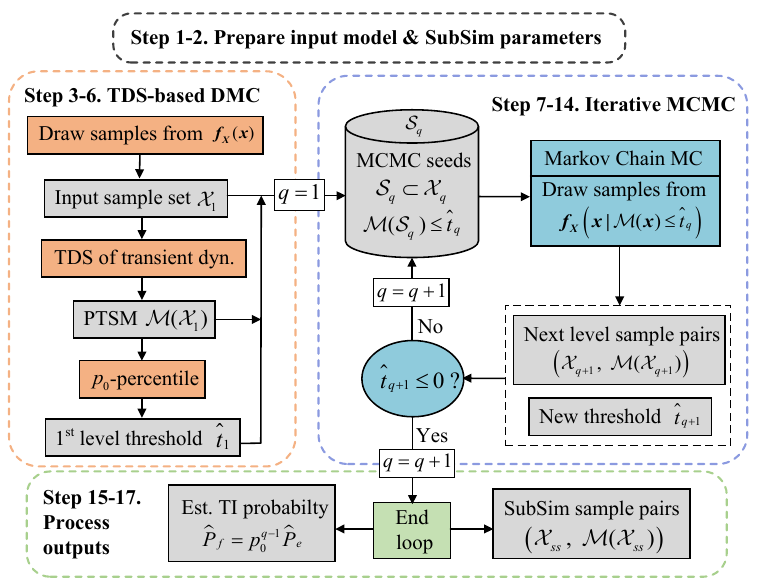}
\vspace{-0.7cm}
\caption{Illustration of the SubSim-based small TI probability estimation process. The details for each step can be found in Algorithm \ref{alg:SubSim_algorithmn}.}
\label{fig:SubSim_explain}
\vspace{-0.1cm}
\end{figure}

Fig. \ref{fig:SubSim_explain} illustrates the SubSim-based small TI probability estimation process 
while the details for each step are summarized in Algorithm \ref{alg:SubSim_algorithmn}. 
\vspace{-0.3cm}

\noindent \textbf{Remarks on the SubSim Parameters (Step \ref{algstep:SS_parameters}):} The values of $p_0$ and $N_{\mathrm{bat}}$ are preferred if both $p_0 N_{\mathrm{bat}}$ and $1/p_0$ are positive integers. Specially, $p_0=0.1$ is suggested by \cite{au2014SubSimBook} to achieve good estimation efficiency, while $N_{\mathrm{bat}}$ should be sufficiently large (e.g., $N_{\mathrm{bat}}\geq 10^3$) to ensure adequate accuracy in both the initial DMC step and subsequent MCMC steps.

\noindent \textbf{Remarks on the MCMC (Step \ref{algstep:MCMC_start} - \ref{algstep:MCMC_end}):} In this paper, the MCMC starts from a set $\mathcal{S}_k$ of $p_0 N_{\mathrm{bat}}$ initial seeds. From each seed, MCMC will generate a Markov chain with $1/p_0-1$ states, ensuring $\mathcal{X}_{k+1}$ in Step \ref{algstep:MCMC_end} consistently contains $p_0 N_{\mathrm{bat}}+(1/p_0-1)p_0 N_{\mathrm{bat}}=N_{\mathrm{bat}}$ samples. Moreover, the component-wise M-H algorithm \cite{papaioannou2015mcmc} is employed for efficient multi-dimension MCMC. 


\begingroup
\renewcommand{\baselinestretch}{1.2}
\begin{algorithm}
\caption{SubSim-based small TI probability estimation}
\begin{algorithmic}[1] \small  
   \State \textbf{Input Model:} 
   The power system transient dynamic model \eqref{eq:Stochastic-PSDAE}; The joint PDF $f_{\bm{X}}(\bm{x})$ of uncertain parameters $\bm{X}$.   
   \State \textbf{Input Parameters for SubSim:} $p_0$ (intermediate subset probability); $N_{\mathrm{bat}}$ (number of samples per intermediate level) \label{algstep:SS_parameters}.
   \State Draw $N_{\mathrm{bat}}$ i.i.d. samples $\mathcal{X}_1=\{\bm{x}_{1}^{(n)}, n=1, ...,N_{\mathrm{bat}}\}$ from $f_{\bm{X}}(\bm{x})$ and calculate their PTSM values $\mathcal{M}(\mathcal{X}_1)=\{\mathcal{M}(\bm{x}_{1}^{(n)}), n=1, ...,N_{\mathrm{bat}}\}$ by the TDS of \eqref{eq:Stochastic-PSDAE}.
   \State Order $\mathcal{X}_1$ in ascending order according to $\mathcal{M}(\mathcal{X}_1)$.
   \State  Find $p_0$-percentile of $\mathcal{M}(\mathcal{X}_1)$ and denote it by $\hat{t}_1$.
   \State Set $q=1$.
   \While{$\hat{t}_q>0$}
   \State Define $\mathcal{S}_q=\mathcal{X}_q\cap\{\bm{x}: \mathcal{M}(\bm{x})\leq\hat{t}_q\}$ of size $p_0 N_{\mathrm{bat}}$. \label{algstep:MCMC_start}
   \State Use $\mathcal{S}_q$ as the seeds to generate  $(1-p_0 )N_{\mathrm{bat}}$ samples from $f_{\bm{X}}(\bm{x}|\mathcal{M}(\bm{x})\leq\hat{t}_q)$ by MCMC. Denote these samples by $\mathcal{C}_q$.
   \State Set $\mathcal{X}_{q+1}=\{\mathcal{S}_q, \mathcal{C}_q\}$. \label{algstep:MCMC_end}
   \State Order $\mathcal{X}_{q+1}=\{\bm{x}_{q+1}^{(n)}, n=1, ...,N_{\mathrm{bat}}\}$ in ascending order according to their PTSM values $\mathcal{M}(\mathcal{X}_{q+1})$.
   \State  Find $p_0$-percentile of $\mathcal{M}(\mathcal{X}_{q+1})$ and denote it by $\hat{t}_{q+1}$.
   \State $q=q+1$. 
   \EndWhile
   \State Estimate $\mathbb{P}(\mathcal{F}_{q}|\mathcal{M}(\bm{x})\leq\hat{t}_{q-1})$ by \eqref{eq:last_level_prob_est}. Denote it by $\hat{P}_{e}$.
   \State $P_f\approx\hat{P}_f=p_0^{q-1}\hat{P}_{e}$
   \State \textbf{Output:} Estimated TI probability $\hat{P}_f$;The selected input samples during the SubSim process $\mathcal{X}_{ss}=\bigcup_{k=1}^{q-1} (\mathcal{X}_k \setminus \mathcal{S}_k) \cup \mathcal{X}_{q}$; The PTSM values $\mathcal{M}(\mathcal{X}_{ss})$ of $\mathcal{X}_{ss}$.  
   \end{algorithmic}
   
    \label{alg:SubSim_algorithmn}
    \vspace{-0.1cm}
\end{algorithm}
  \endgroup

\vspace{-0.3cm}

\section{Unstbale Sample-Based Sensitivity Analysis} 
\label{sec:Sen}

The SubSim results can help identify the input RVs that significantly impact $P_f$. For mutually independent RVs $\bm{X}$,  one way to quantify this influence is the total variation distance (TVD) $\eta_i$: \cite{tsybakov2009introduction}:
\begin{small}
\begin{equation}
\setlength{\abovedisplayskip}{6pt}
\setlength{\belowdisplayskip}{6pt}
     \eta_i = \frac{1}{2}\mathbb{E}_{X_i}[|1-p_i(x_i|\mathcal{F})/p_i(x_i)|]
\label{eq:Sen_idx}
\end{equation}
\end{small}
where $p_i(x_i)$ is the marginal distribution of $X_i$ and $p_i(x_i|\mathcal{F})$ is the marginal distribution of $X_i$ given system transient instability. Notice that $\eta_i\in[0, 1)$, where values approaching $1$ indicate a significant deviation of $p_i(x_i|\mathcal{F})$ from $p_i(x_i)$, highlighting the critical impact of $X_i$ on the system transient instability probability.

 


Obviously, 
\eqref{eq:Sen_idx} cannot be evaluated analytically. Therefore, we employ the sample pairs from SubSim  $(\mathcal{X}_{ss}, \mathcal{M}(\mathcal{X}_{ss}))$ and a sample-based histogram method for its estimation. 
Let $\mathcal{B}_k, k=0,...,q$ denote the bins formed by the conditional events during SubSim. That is, $\mathcal{B}_0 = \{\bm{x}:\mathcal{M}(\bm{x})> \hat{t}_1\}$; $\mathcal{B}_k = \{\bm{x}:\hat{t}_{k+1}<\mathcal{M}(\bm{x})\leq\hat{t}_k\}$ for $1\leq k\leq q-1$; and $\mathcal{B}_q=\mathcal{F}_q$. Meanwhile, separate the domain of each input RV $X_i$ into $H_{s}$ non-overlapping histogram interval $\mathcal{J}_h^i$, $h=1,\dots,H_{s}$ (e.g., for $X_i\sim U(0, 1)$, the $h$-th intervals can be: $\mathcal{J}_h^i=(\frac{h-1}{H_{s}}, \frac{h}{H_{s}})$). Then, for each of these intervals $\mathcal{J}_h^i$, its probability can be estimated as follows:
 \begin{equation}
 \small
 \setlength{\abovedisplayskip}{4pt}
\setlength{\belowdisplayskip}{4pt}
     \mathbb{P}(X_i\in\mathcal{J}_h^i)=\sum_{k=0}^{q}\mathbb{P}(X_i\in\mathcal{J}_h^i|\mathcal{B}_k)\mathbb{P}(\mathcal{B}_k)\approx\sum_{k=0}^{q}\frac{N_{kh}}{N_k}P_k
\label{eq:IntervalP}
 \end{equation}
 \normalsize
 where $N_k$ represents the number of samples within $\mathcal{X}_{ss}$ that fall into the bin $\mathcal{B}_k$; $N_{kh}$ denotes the number of samples in $\mathcal{B}_k$ whose $x_i$ values also fall into the $h$-th interval $\mathcal{J}_h^i$; $P_k = \mathbb{P}(\mathcal{B}_k)$ represents the bin probability,  which is calculated based on $p_0$ and $\hat{P}_e$ from \eqref{eq:p0} and \eqref{eq:last_level_prob_est}. Besides, when transient instability occurs:
 \begin{small}
 \begin{equation}
 \setlength{\abovedisplayskip}{2pt}
 \begin{aligned}
    \mathbb{P}(X_i\in\mathcal{J}_h^i|\mathcal{F})&=\sum_{k=0}^{q}\mathbb{P}(X_i\in\mathcal{J}_h^i|\mathcal{B}_k\cap\mathcal{F})\mathbb{P}(\mathcal{B}_k|\mathcal{F}) \\
    & \approx \frac{\sum_{k=0}^{q}(N_{khF}/N_k)P_k}{\sum_{k=0}^{q}(N_{kF}/N_k)P_k}
\end{aligned}
\label{eq:IntervalP_F}
 \end{equation}
 \end{small}
Herein, $N_{kF}$ denotes the number of samples that lie in $\mathcal{B}_k\cap\mathcal{F}$. Among these $N_{kF}$ samples, the count of samples whose $x_i$ value falls within the interval $\mathcal{J}_h^i$ is represented by $N_{khF}$. 
Furthermore, we can have an interval-based estimation of \eqref{eq:Sen_idx}:
\begin{small}
\begin{equation}
\setlength{\abovedisplayskip}{4pt}
\setlength{\belowdisplayskip}{4pt}
     \eta_i \approx \hat{\eta}_i=\frac{1}{2}\sum_{h=1}^{H_s}|\mathbb{P}(X_i\in\mathcal{J}_h^i) - \mathbb{P}(X_i\in\mathcal{J}_h^i|\mathcal{F})|  
\label{eq:Sen_idx_est}
\end{equation}
\end{small}
This estimation \eqref{eq:Sen_idx_est} can be 
further extended to bivariate cases when two input RVs exhibit correlations. By dividing the domain of a bivariate joint distribution into intervals $\mathcal{J}_{h_i, h_j}^{i,j}$, the bivariate extension of \eqref{eq:Sen_idx_est} to is formulated as follows:
\begin{small}
\begin{equation}
\setlength{\abovedisplayskip}{0pt}
\setlength{\belowdisplayskip}{2pt}
     \hat{\eta}_{i,j}=\frac{1}{2}\sum_{h_i=1}^{H_s^i}\sum_{h_j=1}^{H_s^j}|\mathbb{P}(\bm{X}_{i,j}\in\mathcal{J}_{h_i, h_j}^{i,j}) - \mathbb{P}(\bm{X}_{i,j}\in\mathcal{J}_{h_i, h_j}^{i,j}|\mathcal{F})| 
    \label{eq:Sen_idx_est_2d}
\end{equation}
\end{small}

\begin{figure}[!]
    \centering
    \includegraphics[width=1\columnwidth]{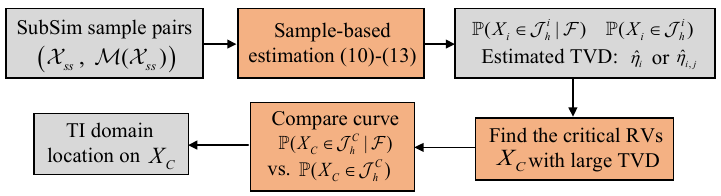}
    \caption{Summary of the proposed sample-based sensitivity analysis method.}
    \label{fig:Sen_disc}
\end{figure}

Finally, notice that while $\eta_i$ serves as a summarized sensitivity index, a comparison of the curve plots for $\mathbb{P}(X_i\in\mathcal{J}_h^i)$ and $\mathbb{P}(X_i\in\mathcal{J}_h^i|\mathcal{F})$ can be used to identify the location of the transient instability domain along the critical input RV $\bm{X}_C$ (i.e., the input RV with large $\eta$ value). Fig. \ref{fig:Sen_disc} summarizes the proposed sample-based sensitivity analysis method.

\section{Case Studies}
\begin{figure}[!h]
\setlength{\abovecaptionskip}{-0.05cm}
\setlength{\belowcaptionskip}{-0.05cm}
    \centering
    \includegraphics[width=0.8\columnwidth]{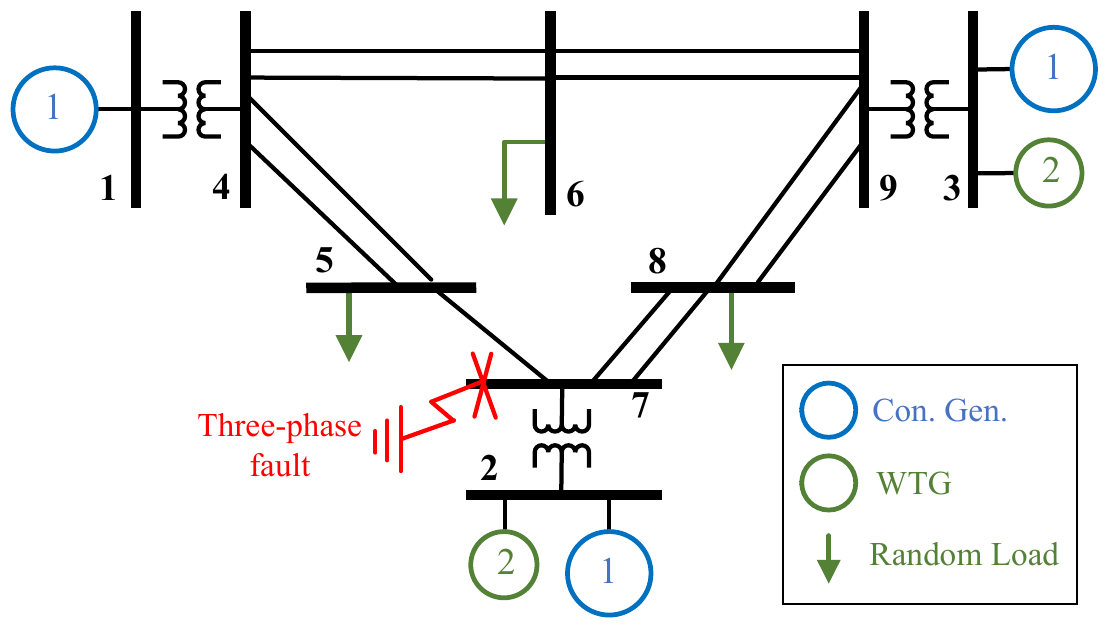}
    \caption{WSCC 3-machine 9-bus system with two wind farms. The blue circles mark the conventional synchronous generators. The green circles indicate the WTGs and the numbers inside the circle represent bus generator indices. The green arrows represent uncertain loads.}
    \label{fig:9bus}
    \vspace{-0.2cm}
\end{figure}

The proposed SubSim-based method is tested on the modified WSCC 3-machine 9-bus test system shown in Fig. \ref{fig:9bus}. 
The DG0S4 synchronous machine, EXC1 exciter and  GOV22 governor in TSAT \cite{powertech_tsat_manual} are used to model the dynamics of the conventional generators. Moreover, both  Bus 2 and Bus 3 are connected to a 50 MW-rated wind farm consisting of GE 1.5 MW wind turbine generators (WTGs) \cite{GE1p5MW_manual}. Specifically, the  wind speed and power output relation of the WTG is described as follows \cite{lu2024advanced}:
\begin{small}
\begin{equation}
    P_w(v) = 
    \begin{cases} 
    P_r \left(\frac{v^3 - v_{in}^3}{v_r^3 - v_{in}^3}\right), & v_{in} \leq v \leq v_r \\
    P_r, & v_r \leq v \leq v_{out} \\
    0, & \text{otherwise}
    \end{cases}
    \label{eq:wind_power_curve}
\end{equation}
\end{small}
 $P_r$, $v_{in}$, $v_{out}$ and $v_r$ denote the rated/max power, cut-in wind speed, cut-out wind speed and rated wind speed of the WTG, respectively. In this paper, we set $v_{in} = 3 $ m/s, $v_{out} = 25$ m/s \cite{GE1p5MW_manual} and  $v_{r} = 11.4$ m/s
\cite{Powertech_User_Manual_2022}. A three-phase ground fault is applied at Bus 7 at $0.1$s and cleared after $T_{\mathrm{fct}} = 6$ cycles by opening Line 5-7. This fault is considered the most severe due to its smallest deterministic CCT among all N-1 contingencies. 

The input uncertainties are modelled by five RVs. In particular, the active power of each of the three constant PQ loads is assumed to follow the Gaussian distribution $\mathcal{N}(1, 0.1^2)$ of its base power \cite{xu2018propagating}.   
The wind speeds at the two wind farms are assumed to follow the Weibull distribution with a scale parameter $11.2$ and a shape parameter $2.2$ \cite{krogsaeter2015validation}. Besides, the three random loads are assumed to vary independently whereas the two wind speed RVs are assumed to be correlated. Their dependencies are modelled by the Gaussian copula with a correlation coefficient $0.8$ \cite{yue2020probabilistic}. No dependency is considered between wind speed and load power.

The TDS-based SubSim and DMC studies of \eqref{eq:Stochastic-PSDAE} are performed using TSAT/DSATools and UQLab \cite{marelli2014uqlab}, in MATLAB R2022b on an Intel Xeon Gold 6126 2.60GHz CPU.

\vspace{-0.3cm}
\subsection{Estimation of Small TI Probability}
As described in Algorithm \ref{alg:SubSim_algorithmn}, we perform SubSim with $p_0=0.1$ and $N_{bat}=10^3$. The Subsim reaches its final $P_f$ estimate after approximately $N=3700$ PTSM sample evaluations \eqref{eq:PTSM}, producing $3700$ sample pairs in $(\mathcal{X}_{ss}, \mathcal{M}(\mathcal{X}_{ss}))$. 

To demonstrate the accuracy of SubSim, we conduct $N_{run}=100$ times of repetitive SubSim runs with the same $p_0$ and $N_{bat}$. From these $100$ estimates of $P_f$, we calculate their average $\Bar{P}_f$ and the estimated coefficient of variation (CoV). For comparison, we also conduct $N_{run}=100$ repetitive DMC runs of $N=3700$ samples (denoted by DMC\_Re), and a single run of benchmark DMC with $N=10^5$ samples (denoted by DMC\_Bmk). The results are listed in Table \ref{tab:Pf_results}.

\begin{table}[h!]
\vspace{-0.3cm}
\renewcommand{\arraystretch}{1.0}
\caption{Comparison of the TI probability prediction accuracy}
\vspace{-0.15cm}
\label{tab:Pf_results}
\centering
\scalebox{1}{
\begin{tabular}{c|c|c||c|c|c}
\hline
Method & $\thead{N}$ & $\thead{N_{run}}$ & $\thead{\Bar{P}_f}$ & $\thead{\hat{\delta}_f}$  & $\thead{\Bar{N}_{TI}}$
\\ [6pt]
\hline
\textbf{SubSim}    & $3700$  &  $100$    & $4.89\times10^{\text{-}4}$ &  $\textbf{32.89\%}$ & $\textbf{489.1}$  \\
\hline
 DMC\_Re    & $3700$  &  $100$    & $4.43\times10^{\text{-}4} $ &  $69.07\%$ & $1.6$  \\
\hline
DMC\_Bmk    & $10^5$  &  $1$    & $4.80\times10^{\text{-}4}$ &  $14.43\%$ & $48$  \\
\hline

\end{tabular}
}
\begin{tablenotes}
\item(1) $N$: total count of PTSM evaluations per run; $N_{run}$: number of repetitive SubSim or DMC runs; $\Bar{P}_f$: average of TI probability estimates of $N_{run}$ runs; $\hat{\delta}_f$: Estimated CoV; $\Bar{N}_{TI}$: average number of TI samples. 
\item(2) $\hat{\delta}_f$ for DMC $10^5$ is by the analytical CoV expression for DMC \cite{au2014SubSimBook}.
\end{tablenotes}
\end{table} 

Table \ref{tab:Pf_results} shows that with the same number of time-consuming PTSM evaluations (about $1.5$ s per evaluation), SubSim achieves much better accuracy on $P_f$ estimation than DMC\_Re ($33\%$ CoV versus $69\%$ CoV). Notably, 22 out of 100 DMC\_Re runs yield a  zero $P_f$ estimate, whereas the lowest $P_f$ estimate by SubSim is $1.75\times10^{\text{-}4}$. This is because SubSim efficiently generates samples in the TI domain (the highest $\Bar{N}_{TI}$ value). This advantage of SubSim can be further exploited in the subsequent sample-based sensitivity analysis.

\vspace{-0.3cm}
\subsection{SubSim Sample-based TI Sensitivity Analysis}

To demonstrate the robustness of the proposed method, all the following TI sensitivity analyses are based on the SubSim run that yielded the lowest $P_f$ estimate. 

\begin{figure}[!ht]
    \centering
    \includegraphics[width=1\columnwidth]{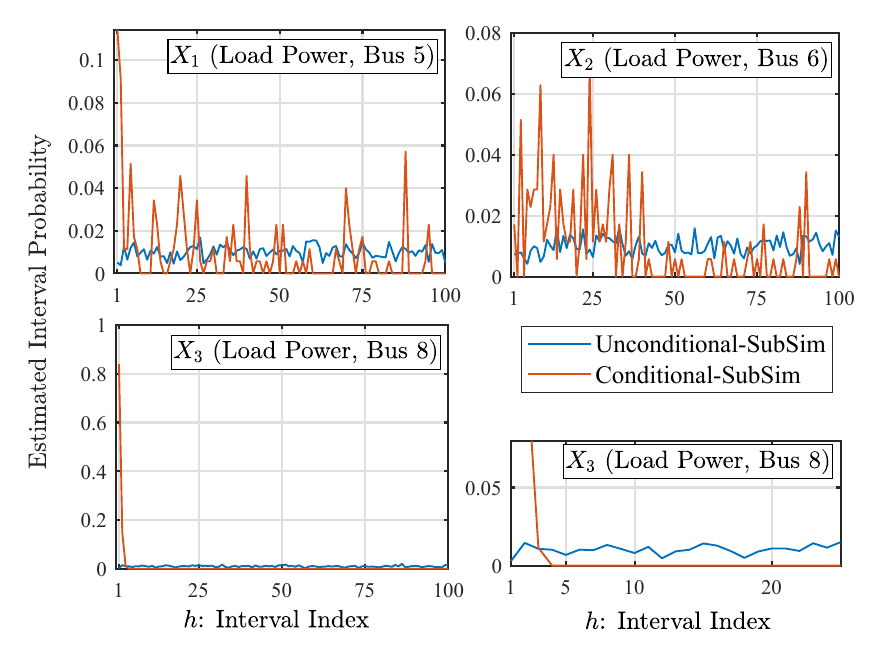}
    \vspace{-0.75cm}
    \caption{Unconditional \eqref{eq:IntervalP} and conditional \eqref{eq:IntervalP_F} marginal interval probability of the independent random inputs estimated by SubSim. 
    The unconditional probabilities are around $1\%$ due to the percentile-based interval division. A conditional probability significantly exceeding $1\%$ implies a potential contribution of the corresponding input to the rare TI events.} 
    \label{fig:sensitivity_1d}
     \vspace{-0.4cm}
\end{figure}

According to Section \ref{sec:Sen}, we start by dividing the input domains into intervals. Specifically, we use the equally spaced percentiles of each input marginal cumulative distribution fuction as the split points. 
The domains of the independent load powers are divided into $100$ intervals each by the $1\%$-percentile, while the domain of the correlated wind speeds is divided into $20^2=400$ intervals by the $5\%$-percentile. Based on this division and the SubSim byproduct $(\mathcal{X}_{ss}, \mathcal{M}(\mathcal{X}_{ss}))$, we calculate the estimated TVD by \eqref{eq:Sen_idx_est} and \eqref{eq:Sen_idx_est_2d}, as listed in Table \ref{tab:ETVD}.
 \vspace{-0.1cm}
\begin{table}[h!]
\vspace{-0.15cm}
    \centering
     \caption{The total variation distance estimated by SubSim}
      \vspace{-0.15cm}  
    \renewcommand{\arraystretch}{1.3}  
    \begin{tabular}{c|c|c|c}
    \hline
    \textbf{${\hat{\eta}_1}$} & \textbf{${\hat{\eta}_2}$} & \textbf{${\hat{\eta}_3}$} & \textbf{${\hat{\eta}_{4,5}}$} \\ \hline
    $0.577$ & $0.556$ & $0.972$ & $0.776$ \\ \hline
    \end{tabular}
    \begin{tablenotes}
    \item * ${\hat{\eta}_1}$-${\hat{\eta}_3}$ are solved by \eqref{eq:Sen_idx_est}. ${\hat{\eta}_{4,5}}$ is solved by \eqref{eq:Sen_idx_est_2d}. ${\hat{\eta}}$ close to $1$ indicates high TI sensitivity.
    \end{tablenotes}
    \label{tab:ETVD}
    \vspace{-0.15cm}
\end{table}
Table \ref{tab:ETVD}  highlights that the $X_3$ (i.e., the load power at Bus 8) has the greatest impact on system TI probability, as indicated by its highest $\hat{\eta}$ value.
Moreover, the comparison between the estimated unconditional \eqref{eq:IntervalP} and conditional \eqref{eq:IntervalP_F} interval probabilities in Fig. \ref{fig:sensitivity_1d}
can provide insights into the location of the TI domain $\mathcal{F}$. 
Significant deviation between \eqref{eq:IntervalP} and \eqref{eq:IntervalP_F} suggests that the corresponding RV might play a critical role in triggering the rare TI events.

In Fig. \ref{fig:sensitivity_1d}, the unconditional interval probabilities (blue curves) remain around $1\%$, validating the SubSim samples-based interval probability estimation \eqref{eq:IntervalP}.
Moreover, Fig. \ref{fig:sensitivity_1d} aligns with our observation from Table \ref{tab:ETVD}: For $X_1$ and $X_2$, their conditional probabilities show no accumulation under a particular operation condition, reflecting low $\hat{\eta}$ values and indicating that $X_1$ and $X_2$ are not the primary contributors to rare TI events; For $X_3$, the high $\hat{\eta}$ value is due to conditional probability accumulation in the extremely low load power region (i.e., $X_3< 82.6$ MW, $4\%$ percentile), which is likely the main cause of rare TI events.


Finally, note that all the above sensitivity analysis is not applicable to DMC\_Re results due to the scarcity of TI samples (i.e., $1.6$ TI samples on average out of $3700$  samples). 

 \vspace{-0.2cm}
\subsection{Elimination of Rare TI Events}
To validate our sensitivity analysis results, compensation tests are performed on the $10^5$ DMC\_Bmk sample pairs, focusing solely on load power for simplicity. Three tests are conducted, each focusing on compensating a single load. For $i=1, 2, 3$, let $\mu_i$ and $\sigma_i$ denote the mean and standard deviation of the random load power $X_i$, respectively. Then, sample with $x_i<\mu_i-1.5\sigma_i$  will be adjusted to $\mu_i-1.5\sigma_i$. 
For example, if we focus on compensating $X_3\sim\mathcal{N}(100, 10^2)$ (the load power at Bus 8), its power consumption will be boosted to at least $85$ MW (e.g., by charging energy storage systems).
Table \ref{tab:TI_events_comp} lists the reductions in TI samples after compensation.
\vspace{-0.4cm}
\begin{table}[h!]
\vspace{-0.2cm}
    \centering
     \caption{The Remaining TI samples after compensation}
      \vspace{-0.15cm}  
    \renewcommand{\arraystretch}{1.4}  
    \begin{tabular}{c|c|c|c|c}
    \hline
     Target RV & Before Comp. & $X_1$ & $X_2$ & \textbf{$X_3$} \\ \hline
    $N_{TI}$  & $48$ & $42$ & $44$ & \textbf{$1$} \\ \hline
    \end{tabular}
    \begin{tablenotes}
        \item * $N_{TI}$: the number of TI samples among $10^5$ DMC samples.
    \end{tablenotes}
    \label{tab:TI_events_comp}
     \vspace{-0.2cm}
\end{table}

\begin{figure}[!ht]
\centering
\includegraphics[width=1\columnwidth]{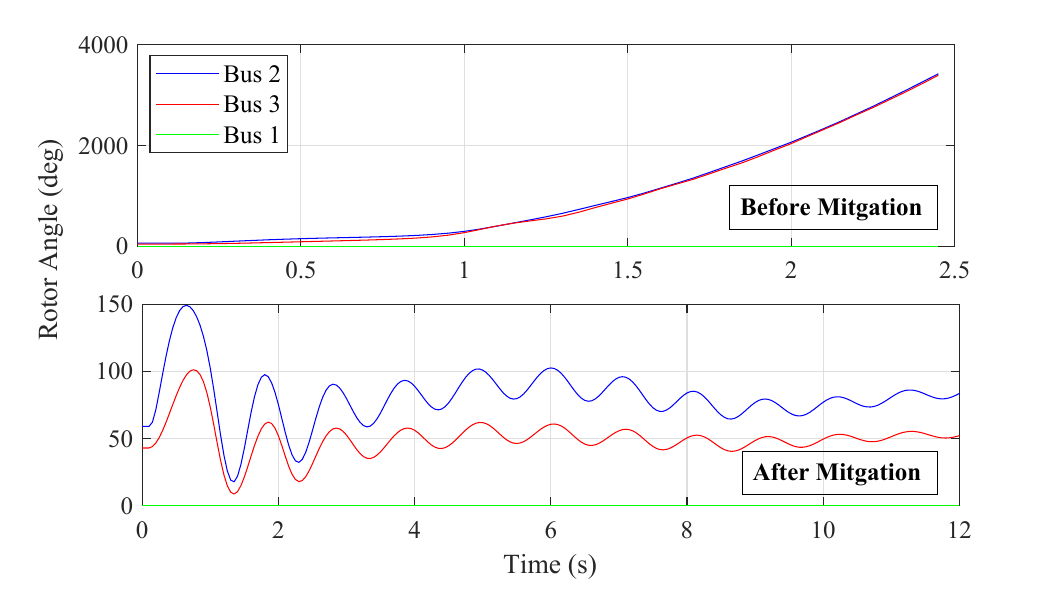}
\vspace{-0.75cm}
\caption{Rotor angle trajectories of one of the $48$ TI samples, before and after the mitigating $X_3$. By boosting $X_3$ (i.e., the load power at Bus 8) from $67.3$ MW to $85$ MW, the CCT increases from $4.5$ cycles to $7 >T_{\mathrm{fct}}=6$ cycles, making the system transient stable.}
\label{fig:Rotor_angle_BA_mitigation}
\end{figure}

The results indicate that boosting $X_3$ 
dramatically reduces the TI probability (i.e., $98\%$ reduction from $4.8\times10^{\text{-}4}$  to $1\times10^{\text{-}5}$) while boosting $X_1$ or $X_2$ yields little improvement (i.e., $13\%$ or $8\%$ reduction). Fig. \ref{fig:Rotor_angle_BA_mitigation} illustrates the stability improvement for one of the $48$ TI samples by boosting $X_3$. 
\vspace{-0.4cm}

In other words, boosting $X_3$ significantly enhances the system's robustness to specific transient faults. Our observations also imply that,  given uncertainties on the generation side, the load side must adapt accordingly to maintain generation-load balance and ensure a sufficient stability margin. 


\section{Conclusions}
This paper proposes a SubSim-based method for small TI probability estimation and rare TI events sensitivity analysis. Simulation results show that the proposed method outperforms DMC in small TI probability estimation by more efficiently sampling the low-probability TI domain. The TI sample-based sensitivity analysis successfully identifies the critical random input and the range of the low-probability TI domain.  Finally, the corresponding compensation strategy significantly reduces the number of rare TI events and enhances the system's robustness to transient faults. 

However, notice that the estimation of $P_f$ is still computationally demanding even with the proposed SubSim-based method. 
Therefore, our future work includes developing PTSA methods with higher efficiency (e.g., surrogate model-based methods)
and applying the proposed method to high-dimension correlated inputs and in large-scale systems.


\bibliographystyle{IEEEtran}
\bibliography{main.bib}

\end{document}